\documentclass[twocolumn,showpacs,preprintnumbers,amsmath,amssymb,
showkeys,floatfix,aps]{revtex4}
\usepackage{graphicx}
\usepackage{dcolumn}
\usepackage{bm}
\usepackage{bigints}
\usepackage{natbib}
\usepackage{amsmath}

\begin{document}

\title{Extraction of the angular power spectrum produced by inflation from
  observations of experiments such as Simons Observatory}

\author{Dmitry I. Novikov\footnote{Corresponding author E-mail: dinovikov27@gmail.com}}
\affiliation{Astro-Space Center of P.N. Lebedev Physical Institute, Profsoyusnaya 84/32, Moscow, Russia 117997}
\author{Kirill O. Parfenov}
\affiliation{Astro-Space Center of P.N. Lebedev Physical Institute, Profsoyusnaya 84/32, Moscow, Russia 117997}
\begin{abstract}

  We demonstrate an approach that allows separating two-point correlations
  created by a Gaussian random field from correlations created by cosmic
  foregrounds such as polarized dust emission, gravitational lensing
  and other non-Gaussian signals. The result of traditional approaches
  should typically be a 'foreground-cleaned' two-dimensional CMB map of
  the anisotropy or polarization. Our method does not create a clean map,
  but extracts the part of the two-point correlations, or equivalently the
  part of the power spectrum, which is due only to the Gaussian component
  of the observed signal produced by inflation.

\end{abstract}


\maketitle

\section{Introduction}

Detection of the primordial B-mode of the cosmic microwave background (CMB)
polarization is one of the main goals of modern observational cosmology.
The cosmological B-mode of polarization
can appear due to the tensor perturbations of the spacetime
metric. Thus, the presence of the B component can confirm the existence of
primordial gravitational waves and the inflationary theory of the evolution
of our Universe.

Next-generation experiments such as SO \cite{2019JCAP...02..056A}, CMB-S4 \cite{2022ApJ...926...54A} and LiteBIRD
\cite{2023PTEP.2023d2F01L}
aim to study the CMB linear polarization and significantly
reduce the
estimation of the upper limit for tensor-to-scalar ratio up
to $r\le 0.01\div 0.001$.
The solution to this problem inevitably faces obstacles in the form of
polarized foregrounds, such as galactic dust and image distortion by
gravitational lensing. All this leads to the appearance of additional
B polarization, which must be separated from the primordial B-mode. The main
approaches
to cleaning the observed signal from foregrounds of various origins are
multi-frequency analysis of CMB maps \cite{1996MNRAS.281.1297T,2017ApJ...842...62S} and delensing \cite{2024PhRvD.110d3532H,2017PhRvD..96l3511Y,2021PhRvD.103b2004B,2022PhRvD.105b3511N}.
However, the lack of our knowledge about the foregrounds spectral properties,
as well as possible inaccuracies in the modeling of gravitational lensing
do not guarantee perfect cleaning and may lead to incorrect interpretation
of the observational data.

We propose a special approach to signal cleaning based on a very unique
property of inflation, namely the Gaussian nature of the initial perturbations
\cite{1982PhLB..117..175S,1997PhRvD..56.1836M}. The results of
WMAP \citep{2013ApJS..208...20B,2013JCAP...07..018C} and Planck
\citep{2014A&A...571A..15P,2020A&A...641A...4P,2020A&A...641A...7P}
experiments confirmed that the observed anisotropy is  very close to Gaussian.
Inflationary theory predicts that two-dimensional maps of the
CMB anisotropy and polarization is formed as a random Gaussian field. This
means that for a clean signal uncontaminated by foregrounds and
undistorted by lensing, the two-point correlations should correspond to the
distribution determined by such a field. Indeed, a random variable in the
form of a product of two values of a Gaussian field separated by a fixed
angular distance must obey the corresponding probability distribution
function, which can be found analytically. Any deviation from such a
distribution means the presence of a non-Gaussian signal in the observational
data. Our task in studying the distribution of two-point correlations is to
separate the Gaussian part of correlations from the non-Gaussian part.
In this way, we reconstruct only that part of the angular power spectrum
that corresponds to the signal generated during inflation.

Note that the result of our approach will not be a cleaned map of polarization
or anisotropy. The outcome of such a method is the angular power spectrum
cleaned of foregrounds and lensing.

The  outline of this paper is as follows:
In Section II we describe our proposed method and in Section
III we provide brief conclusions.

\section{Probability distribution of two-point correlations and the method for extracting the 'clean' power spectrum}

The observed sky map of the anisotropy or polarization of the relic radiation contains the cosmological Gaussian signal of interest to us
$g(\boldsymbol{\eta})$ and foregrounds of various origins
$f(\boldsymbol{\eta})$:
\begin{equation}
s(\boldsymbol{\eta})=g(\boldsymbol{\eta})+f(\boldsymbol{\eta}),
\end{equation}
where $s$ is the total observed signal and $\boldsymbol{\eta}$ is a unit
vector with spherical coordinates $\theta,\varphi$. One can consider 
signal $s$ as either the scalar value T of CMB temperature or the Stokes
parameters Q or U. Two-point correlations $x_r=s_1\cdot s_2$ consist of a set
of products of two observed signals $s_1=s(\boldsymbol{\eta_1})$ and $s_2=s(\boldsymbol{\eta_2})$ separated by a fixed angular distance $r$, where
$\cos(r)=\boldsymbol{\eta_1}\boldsymbol{\eta_1}$. Therefore, one can write:
\begin{equation}
  \begin{array}{l}
    x_r=s_1s_2=g_1g_2+g_1f_2+g_2f_1+f_1f_2.    
    \end{array}
\end{equation}
Let us denote 3 random values as follows:
\begin{equation}
  \begin{array}{l}
    a_r=g_1g_2,\hspace{0.1cm}b_r=g_1f_2+g_2f_1,\hspace{0.1cm}c_r=f_1f_2.
    \end{array}
\end{equation}
The first term in Eq. (2) represents
the correlations of the Gaussian quantities $g_1$ and $g_2$ with zero mean and
variances $\langle g_1^2\rangle=\langle g_1^2\rangle=\xi_0$,
$\langle g_1g_2\rangle=\xi_r$. The probability distribution function
$P_{gg}(a_r)$ for this term can be found analytically.
The second term is the cross product of the Gaussian signal by the
foreground and we denote the PDF for this value as
$P_{gf}(b_r)$. The third term is the
two-point correlations of the foregrounds with distribution $P_{ff}(c_r)$.
Since we assume that the Gaussian
cosmological signal and the foreground are independent, the joint probability
distribution function of all three values $a_r,b_r,c_r$ is the product of the
individual probabilities: $P(a_r,b_r,c_r)=P_{gg}(a_r)P_{gf}(b_r)P_{ff}(c_r)$.

It is convenient to use the following variables:
\begin{equation}
  \begin{array}{l}
    x_r=a_r+b_r+c_r,\hspace{0.1cm}y_r=a_r-b_r+c_r,\\
    z_r=a_r-b_r-c_r.
    \end{array}
\end{equation}
The distribution of the observed two-point correlations $x_r$ is as
follows:
\begin{equation}
  \begin{array}{l}
    P(x_r)=
    \int P_{gg}(\frac{x_r+z_r}{2})P_{gf}(\frac{x_r-y_r}{2})
    P_{ff}(\frac{y_r-z_r}{2})\frac{dy_rdz_r}{4},
    \end{array}
\end{equation}
where the factor $1/4$ appears from the change of variables in Eq. (4).
It is easy to see that for a negligibly small non-Gaussian foreground $f$
all three variables tend to a correlation of two Gaussian values
$a=g_1g_2$: $x_r\rightarrow a$, 
$y_r\rightarrow a$, $z_r\rightarrow a$ and functions
$P_{gf}(\frac{x_r-y_r}{2})$, $P_{ff}(\frac{y_r-z_r}{2})$ actually turn into
delta functions. Thus, the distribution of observed correlations becomes
close to the distribution of Gaussian two-point correlations:
$P(x_r)\rightarrow P_{gg}(a)$.

Thus, the observed distribution of two-point correlations is the distribution
of correlations corresponding to a Gaussian initial signal generated by
inflation plus some deviation $\Delta$:
\begin{equation}
  \begin{array}{l}
    \vspace{0.2cm}
    P(x_r)=P_{gg}(x_r)+\Delta(x_r),\\
    \Delta(x_r)=P(x_r)-P_{gg}(x_r).\\
    \end{array}
\end{equation}

As for the distribution function $P_{gg}(x_r)$, it is easy to find it
analytically. Indeed, the joint probability distribution function of two
Gaussian values $g_1, g_2$ looks as follows:
\begin{equation}
  \begin{array}{l}
    \vspace{0.1cm}
    dP(g_1,g_2)=\frac{1}{4\pi\sqrt{1-\gamma^2}}e^{-\frac{(g_1+g_2)^2}{4(1+\gamma)}}
    e^{-\frac{(g_1-g_2)^2}{4(1-\gamma)}}dg_1dg_2,\\
    \gamma=\frac{\xi_r}{\xi_0}.
    \end{array}
\end{equation}
After some fairly simple algebra we find the distribution for $x_r=g_1g_2$
Fig 1:
\begin{equation}
  \begin{array}{l}
    \vspace{0.1cm}
    P_{gg}(x_r)=\frac{1}{16\pi\sqrt{1-\gamma^2}}
    \int\limits_1^\infty e^{-\frac{(t-\gamma)x_r}{4(1-\gamma^2)}}
    \frac{dt}{\sqrt{t^2-1}}.
    \end{array}
\end{equation}
The shape of this distribution is determined by a single parameter
$\xi_r$. Therefore one can write $P_{gg}(x_r)=P_{gg}(x_r,\xi_r)$ and obviously:
\begin{equation}
  \begin{array}{l}
    \vspace{0.1cm}
    \int\limits_{-\infty}^{\infty}P_{gg}(x_r)x_rdx_r=\xi_r
    \end{array}
  \end{equation}

The total number of two-point correlations $x_r$ on a pixelized sky map can
be estimated as $N\cdot r/h$, where $N$ is the total number of pixels and $h$ is
the pixel size. Thus, at good resolution a detailed distribution
function of the observed value $x_r$ can be constructed.

In order to extract  the component of the power spectrum, determined by the
initial Gaussian field, from the observed signal T, E or B it is
necessary to construct the distribution $P(x_r)$ of the observed two-point
correlations $x_r$ for each angular distance r between the two signals in
the sky. Correct fitting of the observed distribution $P(x_r)$ by the function
$P_{gg}(x_r,\xi_r)$ will give us the correct value of the two-point
correlation function $\xi_r$. The equivalent power spectrum is obtained by
decomposing $\xi_r$ into Legendre polynomials.

To correctly fit the observed distribution by the parameter
$\xi_r$, one can use an
approach similar to the Internal Linear Combination (ILC) method
\cite{1992ApJ...398..169R}, in which
the additional non-Gaussian component $\Delta$ in the Eq. 6 should be
interpreted as noise.

\begin{figure}[tbh]
  \includegraphics[width=1\columnwidth]{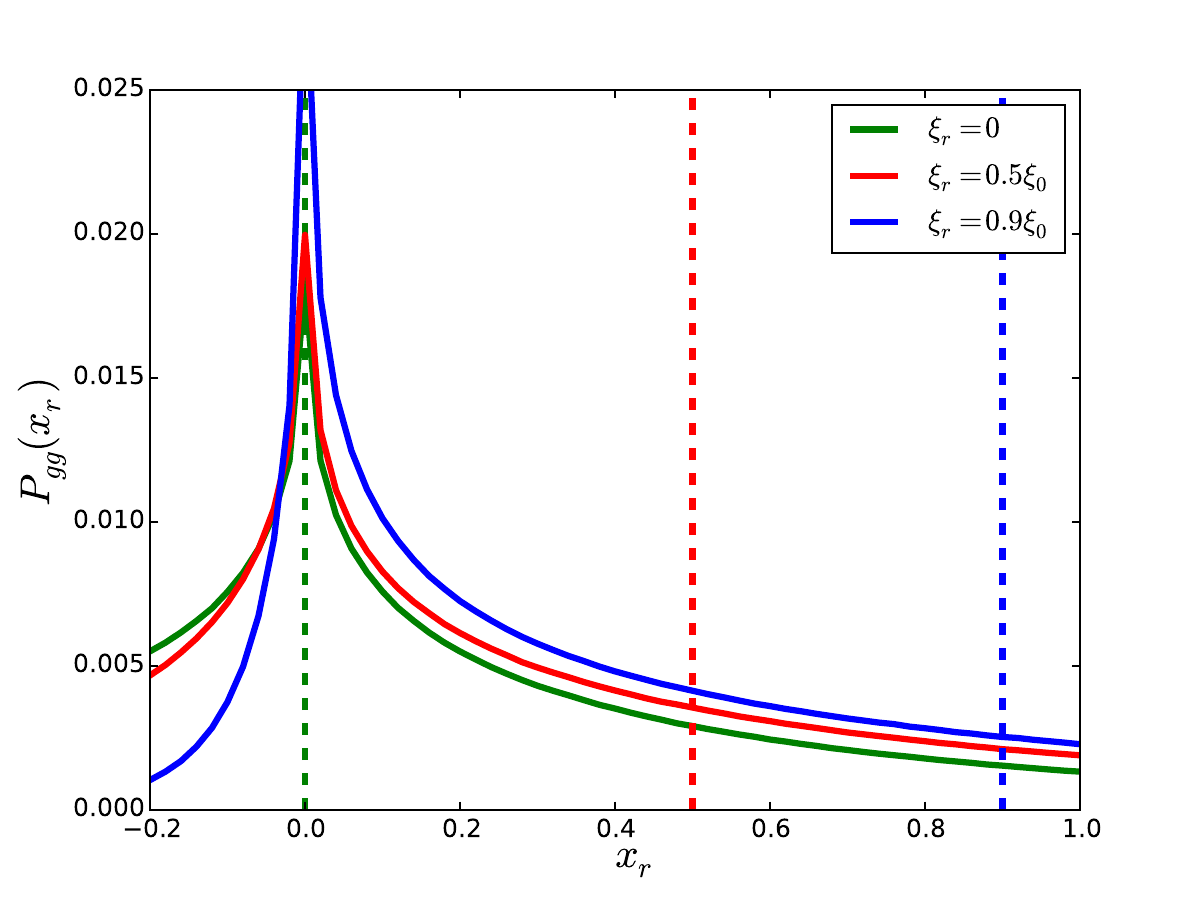}
  \caption{Probability distribution functions for two-point correlations
    of a Gaussian signal. The dashed vertical lines indicate the average
    values of the correlations, equal to the values of the correlation
    functions.}
\end{figure}

\section{Conclusions}
In this paper we propose a method for extracting
the part of the power spectrum
that corresponds only to the Gaussian component of the signal in the
observational
data of CMB anisotropy or polarization.
With good statistics, this can help to extract the spectrum of the
cosmological B mode of polarization, corresponding to the
primordial gravitational waves.

Our approach is based on the Gaussian nature of the perturbations generated
during inflation and can serve as a good complement to the methods used to
clean the signal from foregrounds. Unlike traditional methods, we do not
reconstruct the CMB map, but we propose to extract the correct part of the
power spectrum from observational data.

The approach we propose can be modified and improved by using additional
information about two-point correlations of foregrounds and the correlation properties of the picture produced by the gravitational lensing.
In this case, when
analyzing the distributions of two-point correlations, it makes sense to use analogues of the modified ILC method, such as
cILC \cite{2011MNRAS.418..467R,2020MNRAS.494.5734R,2013PhRvD..88f3526H},
MILC \cite{2005MNRAS.357..145S,2017MNRAS.472.1195C,2021MNRAS.500..976R}
or LRM \cite{2024PhRvD.109b3523M}.

\newpage
The work is supported by the Project No. 36-2024 of LPI new scientific
groups.

\def\apj{Astrophys.~J}
\def\apjl{Astrophys.~J.,~Lett}
\def\apjs{Astrophys.~J.,~Supplement}
\def\an{Astron.~Nachr}     
\def\aap{Astron.~Astrophys}
\def\mnras{Mon.~Not.~R.~Astron.~Soc}
\def\pasp{Publ.~Astron.~Soc.~Pac}
\def\aaps{Astron.~and Astrophys.,~Suppl.~Ser}
\def\apss{Astrophys.~Space.~Sci}
\def\ibvs{Inf.~Bull.~Variable~Stars}
\def\japa{J.~Astrophys.~Astron}
\def\na{New~Astron}
\def\aspproc{Proc.~ASP~conf.~ser.}
\def\aspcs{ASP~Conf.~Ser}
\def\aj{Astron.~J}
\def\actaa{Acta Astron}
\def\araa{Ann.~Rev.~Astron.~Astrophys}
\def\caosp{Contrib.~Astron.~Obs.~Skalnat{\'e}~Pleso}
\def\pasj{Publ.~Astron.~Soc.~Jpn}
\def\memsai{Mem.~Soc.~Astron.~Ital}
\def\astl{Astron.~Letters}
\def\aipproc{Proc.~AIP~conf.~ser.}
\def\physrep{Physics Reports}
\def\jcap{Journal of Cosmology and Astroparticle Physics}
\def\baas{Bulletin of the AAS}
\def\ssr{Space~Sci.~Rev.}
\def\azh{Astronomicheskii Zhurnal}

\bibliography{aaaso.bib}




\end{document}